\documentclass[twocolumn,aps,showpacs,superscriptaddress,longbibliography]{revtex4-1}
\usepackage{graphicx}
\usepackage{latexsym}

\begin{document}

\title{Experimental quantification of the entanglement of noisy twin beams}

\author{V\'{a}clav Mich\'{a}lek}
\email{vaclav.michalek@upol.cz} \affiliation{Joint Laboratory of Optics of
Palack\'{y} University and Institute of Physics of CAS, Faculty of Science,
Palack\'{y} University, 17. listopadu 50a, 771 46 Olomouc, Czech Republic}
\author{Jan Pe\v{r}ina,~Jr.}
\email{jan.perina.jr@upol.cz} \affiliation{Joint Laboratory of Optics of
Palack\'{y} University and Institute of Physics of CAS, Faculty of Science,
Palack\'{y} University, 17. listopadu 50a, 771 46 Olomouc, Czech Republic}
\author{Ond\v{r}ej Haderka}
\affiliation{Institute of Physics of Academy of Sciences of the Czech Republic,
Joint Laboratory of Optics of Palack\'{y} University and Institute of Physics
of AS CR, 17. listopadu 12, 772 07 Olomouc, Czech Republic}

\begin{abstract}
Gradual loss of the entanglement of a twin beam containing around 25 photon
pairs with the increasing external noise is experimentally investigated. The
entanglement is quantified by the non-classicality depths and the
non-classicality counting parameters related to several non-classicality
criteria. The reduction of intensity moments of the analyzed multi-mode twin
beams to single-mode ones allows to determine the negativity as another
quantifier of the entanglement. Both the raw photocount histograms and the
reconstructed photon-number distributions are analyzed in parallel.
\end{abstract}


\maketitle

\section{Introduction}

Twin beams (TWBs) ideally composed of photon pairs have very interesting
quantum properties: They exhibit the entanglement between the photons belonging
to the same photon pair that occurs in different degrees of freedom including
frequencies, polarizations or propagation directions. At the same time,
however, the TWBs containing on average typically more than one photon pair
exhibit perfect correlations between the numbers of the signal and idler
photons, that represent another attribute of the TWB quantumness. The
entanglement in the TWB, as the TWB prominent feature, finds its applications
in metrology (measurement of ultra-short time intervals, absolute detector
calibration \cite{Migdall1999,PerinaJr2012a}), quantum communications
(reduction of noise, quantum cryptography) and various quantum-information
protocols~\cite{Nielsen2000}. Quantum states with specific properties may be
obtained using various types of post-selection realized on the TWB
\cite{MaganaLoaiza2019}. However, the noise superimposed on the TWB occurs in a
smaller or greater amount in all these applications. For example, in the
quantum-communication applications the noise increases linearly with the
distance~\cite{Allevi2018}. As certain minimal amount of the entanglement is
indispensable for all applications of TWBs, restriction to the maximal
tolerable amount of the noise occurs. This brings the need to quantify the TWB
entanglement and its relationship to the noise. The noise may originate either
in the sources outside the TWB or in photon pairs of the TWB being partly
absorbed during their propagation (typically in optical fibers). In this
contribution, we suggest three theoretical concepts how to quantify the TWB
entanglement. We verify these concepts experimentally: We generate a TWB with
around 25 photon pairs on average and superimpose an additional noise with the
increasing intensity onto both signal and idler beams.

Quantification of the entanglement of TWBs is not an easy task because the TWBs
are typically (spectrally and spatially) multi-mode and as such they are
properly characterized by quasi-distributions of the overall signal and idler
(integrated) intensities, instead of amplitudes. This comes from the fact that
the multi-mode character of the fields makes the information about the phases
of individual spatio-spectral modes as well as their individual intensities
unimportant. A larger number of modes prevents the application of the homodyne
tomography \cite{Leonhardt1997,Lvovsky2009} in the experimental investigations
of TWBs, as well as the use of the entanglement witnesses based on the moments
of fields' amplitudes \cite{Duan2000,Simon2000,Shchukin2005,Miranowicz2010}.
Quantification of the entanglement of multi-mode optical fields represents a
serious and demanding problem even in specific cases when individual modes and
their inter-modal correlations are measured \cite{Gerke2015,Harder2016}. In the
case of multi-mode TWBs, we do not have access to the properties of individual
modes. However, we know that the reduced states of the signal and idler beams
are multi-mode thermal \cite{Mandel1995}, i.e. they are purely classical, as a
consequence of the spontaneous emission of photon pairs in the process of
spontaneous parametric down-conversion \cite{Boyd2003}. This means that the
quantification of TWB entanglement can be mapped onto the quantification of the
TWB non-classicality.

In general, the non-classicality of a state is recognized by the negative
values of quasi-distributions of intensities (even being in the form of
generalized functions) \cite{Glauber1963,Perina1991}. In the case of multi-mode
TWBs, the problem of non-classicality identification can be considerably
simplified when applying suitable non-classicality identifiers/witnesses
(NI)~\cite{PerinaJr2017a,Arkhipov2018a,Arkhipov2016c} that are conveniently
based on the intensity moments. The fields' intensities and their moments can
be measured by photon-number-resolving detectors that provide the corresponding
photocount distributions
\cite{Haderka2005,Allevi2012,PerinaJr2013a,Harder2016,Chesi2019,MaganaLoaiza2019}.
We note that also the NIs based directly on the elements of photocount (or
photon-number) distributions may also be used for this
purpose~\cite{Klyshko1996,Waks2006,Wakui2014,PerinaJr2017a}. The quantification
of non-classicality/entanglement is then reached by applying the concept of
the Lee non-classicality depth \cite{Lee1991} or the approach leading to the
non-classicality counting parameter \cite{PerinaJr2019}.

Here, we suggest and verify an alternative approach in which we first determine
the intensity moments appropriate to one typical (paired) mode and then we use
these intensity moments in the formula for the negativity of a Gaussian
two-mode field \cite{Adesso2007,Arkhipov2015} to directly quantify the TWB
entanglement. The negativity \cite{Hill1997} exploits the properties of the
partially-transposed statistical operator \cite{Peres1996,Horodecki1996} to
quantify the amount of the entanglement in a composed quantum system.

The paper is organized as follows. Non-classicality and entanglement
identifiers and quantifiers are theoretically introduced in Sec.~II. The
experimental setup, performed experiment and the reconstruction method for
revealing a TWB joint photon-number distribution from the experimental
photocount histogram are described in Sec.~III. Degradation of the
non-classicality and entanglement caused by an additional noise with the
increasing intensity is discussed in Sec.~IV using the theoretical tools of
Sec.~II. Sec.~V brings conclusions.

\section{Non-classicality and entanglement identification and quantification}

For TWBs, the noise-reduction-factor $ R $ is the commonly determined quantity
that may also indicate their non-classicality:
\begin{equation}  
 R = 1 + \frac{ \langle \left[\Delta (W_{\rm s}-W_{\rm i})\right]^2\rangle }{ \langle{W_{\rm s}}\rangle + \langle{W_{\rm i}}\rangle },
\label{1}
\end{equation}
where $ W_{\rm s} $ ($ W_{\rm i} $) denotes the signal- (idler-) field
(integrated) intensity and $ \Delta W = W - \langle W\rangle $. According to
its definition the noise-reduction-factor $ R $ quantifies pairing of the
photons in a TWB. For an ideal TWB composed of only photon pairs, it equals to
zero. If an additional noise on the top of the paired photons is present in the
TWB, $ R > 0 $. The larger the amount of the noise, the greater the value of $
R $. It can be shown that the TWBs with $ R < 1 $ are nonclassical.

The intensity moments \cite{Mandel1995,Perina1991} needed for the determination
of the noise-reduction-factor $ R $ as well as other characteristics of the
TWBs are commonly derived from the moments of the reconstructed photon-number
distribution $ p(n_{\rm s},n_{\rm i}) $. This distribution is obtained by the
reconstruction from the experimental photocount histogram $ f(c_{\rm s},c_{\rm
i}) $. The intensity moments $ \langle W_{\rm s}^k W_{\rm i}^l\rangle $
represent the normally-ordered photon-number moments. They are derived from the
usual photon-number moments $ \langle n_{\rm s}^i n_{\rm i}^j \rangle $ using
the following linear relations valid for one effective bosonic mode with the
operators fulfilling the canonical commutation relations ($ k,l =1,2,\ldots $)
\cite{Saleh1978,Mandel1995,Perina1991}:
\begin{eqnarray}  
 \langle W_{\rm s}^k W_{\rm i}^l\rangle = \sum_{m=0}^k
   S(k,m) \sum_{j=0}^{l}S(l,j) \langle n_{\rm s}^m n_{\rm i}^j \rangle .
\label{2}
\end{eqnarray}
In Eq.~(\ref{2}), symbol $ S $ stands for the Stirling numbers of the first
kind \cite{Gradshtein2000}.

The reconstruction of a photon-number distribution removes the 'distortions' in
the experimental photocount histogram caused by the detector. As such it
improves in general the characteristics of the analyzed field, especially its
non-classicality. To assess the parameters/quality of the directly measured
photocount histogram, we may assume that it was obtained by an ideal detector
whose operation does not require any correction. In this case, we may consider
in the r.h.s. of Eq.~(\ref{2}) the photoucount moments $ \langle c_{\rm s}^i
c_{\rm i}^j\rangle $ instead of the photon-number moments $ \langle n_{\rm s}^i
n_{\rm i}^j\rangle $ and determine the corresponding intensity moments. Such
intensity moments derived from the photocount moments can then be used in
parallel to the usual intensity moments of Eq.~(\ref{2}) to determine the
quantities of interest and discuss the related properties. We note that we
systematically use the quantities $ c_{\rm s} $ and $ c_{\rm i} $ to count the
numbers of detected electrons (photocounts) whereas the numbers $ n_{\rm s} $
and $ n_{\rm i} $ quantify photon numbers in the reconstructed TWB.

The real experimental quantification of the TWB non-classicality can be based
upon suitable NIs for which the non-classicality depths $ \tau $ introduced in
\cite{Lee1991} or the non-classicality counting parameters $ \nu $ defined in
\cite{PerinaJr2019} are determined (for details, see below). Following the
comprehensive analysis of NIs based on the intensity moments of TWBs
\cite{PerinaJr2017a}, we consider the following three representative NIs:
\begin{eqnarray}   
 M &\equiv& \langle W_{\rm s}^2 \rangle \langle W_{\rm i}^2\rangle- \langle W_{\rm s} W_{\rm i}\rangle^2<0, \nonumber \\
 E_2 &\equiv& \langle W_{\rm s}^2\rangle + \langle W_{\rm i}^2 \rangle -2 \langle W_{\rm s}
  W_{\rm i}\rangle < 0 , \nonumber \\
 E_3 &\equiv& \langle W_{\rm s}^3\rangle + \langle W_{\rm i}^3 \rangle - \langle W_{\rm s}^2
  W_{\rm i}\rangle - \langle W_{\rm s} W_{\rm i}^2\rangle <0. 
\label{3}
\end{eqnarray}
The NI $ M $ has a privileged position among other NIs based on the intensity
moments as it only identifies the non-classicality in an arbitrary single-mode
TWB \cite{Arkhipov2018a}. Whereas the NI $ M $ contains the intensity moments
in the cumulative fourth order, the other considered NI $ E_2 $ uses just the
second-order intensity moments. For this reason, the  most commonly applied NI
$ E_2 $ is determined with better experimental precision than the NI $ M $. We
note that for a balanced TWB with $ \langle W_{\rm s}\rangle = \langle W_{\rm
i}\rangle $, $ E_2 < 0 $ is equivalent to $ R < 1 $. In general, the condition
$ R < 1 $ can be transformed into the inequality
\begin{equation} 
 E_2 + \left( \langle W_{\rm s}\rangle - \langle W_{\rm i}\rangle \right)^2 <0
\label{4}
\end{equation}
and so the NI $ E_2 $ is stronger in identifying the non-classicality than the
noise-reduction-factor $ R $. On the other hand, the last considered NI $ E_3 $
directly involves the third-order intensity moments and as such it monitors the
higher (third) -order non-classicality.

The performance of the above NIs can directly be compared for single-mode
fields. In this case, a TWB is nonclassical provided that $ Q \equiv 2\langle
W_{\rm s}\rangle \langle W_{\rm i}\rangle - \langle W_{\rm s}W_{\rm i}\rangle <
0 $ \cite{Perina2011}. Using the formulas $ \langle W_a^2\rangle = 2 \langle
W_a\rangle^2 $, $ \langle W_a^3\rangle = 6 \langle W_a\rangle^3 $, $ a={\rm
s,i} $, $ \langle W_{\rm s}^2W_{\rm i}\rangle = 2 \langle W_{\rm s}W_{\rm
i}\rangle \langle W_{\rm s}\rangle $, and $ \langle W_{\rm s}W_{\rm i}^2\rangle
= 2 \langle W_{\rm s}W_{\rm i}\rangle \langle W_{\rm i}\rangle $ valid for the
single-mode Gaussian fields, we rewrite Eqs.~(\ref{3}) in the form:
\begin{eqnarray} 
 M &=& Q (2\langle W_{\rm s}\rangle \langle W_{\rm i}\rangle + \langle W_{\rm s} W_{\rm i}\rangle) <0, \nonumber \\
 E_2 &=& 2Q + 2( \langle W_{\rm s}\rangle - \langle W_{\rm i}\rangle)^2 < 0 , \nonumber \\
 E_3 &=& 2Q(\langle W_{\rm s}\rangle + \langle W_{\rm i}\rangle) + 2( \langle W_{\rm s}\rangle^3 +
  \langle W_{\rm i}\rangle^3) \nonumber \\
 & & \mbox{} + 4( \langle W_{\rm s}\rangle - \langle W_{\rm i}\rangle)^2
  ( \langle W_{\rm s}\rangle + \langle W_{\rm i}\rangle) <0. 
\label{5}
\end{eqnarray}
According to Eqs.~(\ref{5}), the NI $ M $ identifies all nonclassical
single-mode TWBs, whereas the NIs $ E_2 $ and $ E_3 $ are weaker than the
condition $ Q<0 $. We note that nonclassical balanced TWBs are also completely
identified by the NI $ E_2 $.

The concept of the non-classicality depth (ND) $ \tau $ \cite{Lee1991} is based
upon the behavior of quasi-distributions in the phase space of an optical field
in relation to different field-operator orderings. It uses the fact that the
amount of non-classicality decreases as we move from the normal ordering, that
corresponds to the usual detection by quadratic intensity detectors, to the
anti-normal ordering, in which any optical field exhibits only the classical
properties. The ND $ \tau $ gives the distance on the ordering-parameter axis $
s $ between the point at which the non-classicality is lost $ s_{\rm th} $ and
the point of the normal ordering $ s=1 $:
\begin{equation}    
 \tau = (1-s_{\rm th} )/ 2.
\label{6}
\end{equation}
The threshold ordering parameter $ s_{\rm th} $ is determined so that the
corresponding $ s $-ordered intensity moments $ \langle W_{\rm s}^k W_{\rm
i}^l\rangle_s $ nullify the corresponding NI. The $ s $-ordered intensity
moments are given as~\cite{Perina1991}:
\begin{equation}   
 \langle W_{\rm s}^k W_{\rm i}^l \rangle_s = \left(\frac{2}{1-s}\right)^{k+l} \left\langle
  {\rm L}_k \left(\frac{2W_{\rm s}}{s-1}\right) {\rm L}_l \left(\frac{2W_{\rm i}}{s-1}\right)\right\rangle
\label{7}
\end{equation}
and $ {\rm L}_k $ denotes the $ k $-th Laguerre polynomial
\cite{Gradshtein2000}. Whereas we have $ 0\le \tau \le 1 $ for an arbitrary
field, the value of ND $\tau $ of any nonclassical Gaussian beam cannot exceed
1/2.

On the other hand, the non-classicality counting parameter (NCP) $ \nu \ge 0 $
\cite{PerinaJr2019} is defined as the mean number of photons of a superimposed
(convolved) chaotic field needed to conceal the non-classicality indicated by
the corresponding NI. In this definition the photon-number distribution of the
noisy photons added into the beams is assumed in the form of a single-mode
thermal field which results in the following combined photon-number
distribution $ p^\nu $,
\begin{eqnarray}  
  &p^\nu(n'_{\rm s},n'_{\rm i};\nu) = \sum_{n_{\rm s}=0}^{n'_{\rm s}} \sum_{n_{\rm i}=0}^{n'_{\rm i}} p(n_{\rm s},n_{\rm i})& \nonumber \\
  &\times p^{\rm th}(n'_{\rm s}-n_{\rm s};\nu,1)  p^{\rm th}(n'_{\rm i}-n_{\rm i};\nu,1),&
\label{8}
\end{eqnarray}
that is applied in the above discussed NIs. The photon-number distribution $
p^{\rm th} $ for a $ K $-mode thermal field with $ \langle n \rangle $ mean
photons is given by the Mandel-Rice formula:
\begin{equation}  
 p^{\rm th}(n;\langle n \rangle,K) = \frac{\Gamma(n +K)}{n!\Gamma(K)} \frac{ \langle n \rangle^n }{
  (1 + \langle n \rangle)^{n+K}};
\label{9}
\end{equation}
$ \Gamma $ stands for the gamma function.

Provided that the numbers $ K_{\rm s} $ and $ K_{\rm i} $ of modes in the
signal and idler beams, respectively, are close and are determined by the
formula for a multi-mode thermal field \cite{Perina1991},
\begin{equation} 
 K_a =  \frac{\langle W_a\rangle^2 }{ \langle (\Delta W_a)^2\rangle  }, \hspace{3mm} a ={\rm s,i},
\label{10}
\end{equation}
we may derive single-mode moments $ \langle w_{\rm s}^k w_{\rm i}^l \rangle_s
$. They characterize a typical paired mode and the whole TWB is then considered
as composed of a given number of identical typical paired modes. As the
analyzed TWBs contain several tens of spatio-spectral modes, this approximate
TWB decomposition is well justified. The mean single-mode intensities $ \langle
w_{\rm s}\rangle $ and $ \langle w_{\rm i}\rangle $ are given as:
\begin{equation} 
 \langle w_a\rangle = \frac{\langle W_a\rangle }{K}, \hspace{3mm} a ={\rm s,i},
\label{11}
\end{equation}
where $ K = (K_{\rm s} + K_{\rm i})/2 $ is the average number of modes.
Higher-order single-mode intensity moments are then conveniently derived by
invoking the following relations for the single-mode intensity fluctuations $
\Delta w_{\rm s} $ and $ \Delta w_{\rm i} $:
\begin{equation} 
 \langle (\Delta w_{\rm s})^k (\Delta w_{\rm i})^l \rangle = \frac{ \langle (\Delta W_{\rm s})^k
 (\Delta W_{\rm i})^l \rangle }{K}.
\label{12}
\end{equation}
Using the relations in Eq.~(\ref{12}) the single-mode intensity moments are
determined step by step starting from those for the lowest orders, i.e., from $
\langle w_a^2\rangle $ for $ a ={\rm s,i} $ and $ \langle w_{\rm s} w_{\rm
i}\rangle $.

The single-mode intensity moments then allow us to directly determine the
negativity $ E_{\rm N} $~\cite{Adesso2007,Arkhipov2015}, that is a genuine
entanglement quantifier, along the formula:
\begin{eqnarray}       
 E_{\rm N} &=& \bigl\{ 2b_{\rm p}-(b_{\rm s}+b_{\rm i})(4b_{\rm p}+1)-4b_{\rm s}b_{\rm i}\nonumber \\
 & & \mbox{} +\sqrt{(b_{\rm s}-b_{\rm i})^2+4b_{\rm p}(b_{\rm p}+1)} \bigr\} \nonumber \\
 & & \mbox{} \times \bigl\{
 4(b_{\rm s}+b_{\rm i})(2b_{\rm p}+1)+8b_{\rm s}b_{\rm i}+2\bigr\}^{-1}
\label{13}
\end{eqnarray}
in which $ b_{\rm p} = -1/2 + \sqrt{ 1/4 - \langle \Delta w_{\rm s} \Delta w_{\rm i}) \rangle } $ and $ b_a = \langle
w_a\rangle - b_{\rm p} $ for $ a = {\rm s,i} $. We note that nonzero negativity $ E_{\rm N} $ of an entangled
two-mode beam implies the fulfillment
of the commonly used NIs for such beams \cite{Simon2000,Duan2000,Perina2011}.

\section{Experimental setup and twin-beam reconstruction}

\begin{figure}  
 \centerline{\includegraphics[width=0.92\hsize]{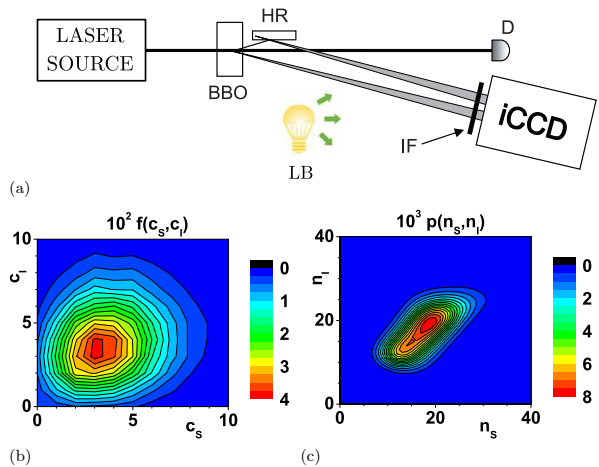}}
 \caption{(a) Scheme of the experimental setup: nonlinear crystal BBO producing
 a TWB; mirror HR reflecting the idler beam; light bulb LB emitting the noisy field
 with defined intensity uniform over the iCCD; bandpass interference filter IF; intensified CCD camera
 iCCD; detector D used for pump-beam stabilization. (b) Normalized experimental photocount histogram
 $f(c_{\rm s},c_{\rm i})$ giving the number of realizations with $ c_{\rm s} $ and $ c_{\rm i} $ registered
 electrons (photocounts) and (c) the corresponding reconstructed photon-number distribution $p(n_{\rm s},n_{\rm i})$
 of the least-noisy TWB.}
\label{fig1}
\end{figure}

In the experiment whose scheme in shown in Fig.~\ref{fig1}(a), a noiseless TWB
was generated in a 5-mm-long type-I $ \beta $-barium-borate crystal (BaB$ {}_2
$O$ {}_4 $, BBO) cut for a slightly non-collinear geometry. Parametric
down-conversion was pumped by pulses originating in the third harmonic (280~nm)
of a femtosecond cavity-dumped Ti:sapphire laser (pulse duration 180~fs at the
central wavelength of 840~nm, repetition rate 50~kHz, pulse energy 20~nJ at the
output of the third harmonic generator). The external noise was produced by a
bulb lamp with variable light intensity. The signal, idler and noise fields
were detected in three different equally-sized detection regions (in the form
of strips) on the photocathode of an iCCD camera Andor DH 345-18U-63. The
camera set for the 4~ns-long detection window was driven by the synchronization
electronic pulses from the laser and it operated at 14~Hz frame rate. Whereas
two detection regions that monitored the signal and idler beams contained both
photons from pairs and the noise photons, the third detection region was
illuminated only by the noise photons thus gave the intensity of the
superimposed noise field. The photons of all three fields impinging on the
camera were filtered by a 14-nm-wide bandpass interference filter with the
central wavelength at 560~nm. As the bandwidth of the spectral intensity
cross-correlation function of the TWB equals around 2~nm under the used
conditions, the edge effects of the filters causing losses of photons from
photon pairs did not have to be explicitly considered. The pump intensity, and
thus also the TWB intensity, was actively stabilized by means of a motorized
half-wave plate followed by a polarizer and a detector that monitored the
actual intensity.

In the experiment, we first investigated the TWB without an additional noise.
The experimental photocount histogram $f(c_{\rm s},c_{\rm i})$ obtained after $
10^4 $ measurement repetitions as well as the reconstructed photon-number
distribution $p(n_{\rm s},n_{\rm i})$ are plotted in Figs.~\ref{fig1}(b) and
\ref{fig1}(c). This TWB caused on average $\langle c\rangle=5.5$ photocounts
per detection region, which corresponds to $ \langle n\rangle=\langle W\rangle
= 24.4 $ photon pairs in a TWB. Owing to non-ideal detection efficiency of the
iCCD camera the joint photocount distribution $ f $ is smeared from the
diagonal given as $ c_{\rm s} = c_{\rm i} $. The reconstruction tends to
eliminate this smearing, but still a typical droplet shape is observed for the
photon-number distribution $ p $. The maximum-likelihood approach was applied
to arrive at the photon-number distribution $ p(n_{\rm s},n_{\rm i}) $ in the
form of a steady state of the following iteration procedure
\cite{Dempster1977,PerinaJr2012} ($ l=0,1,\ldots $):
\begin{eqnarray} 
 p^{(l+1)}(n_{\rm s},n_{\rm i}) &=& p^{(l)}(n_{\rm s},n_{\rm i})
  \nonumber \\
 & & \hspace{-20mm} \mbox{} \times
  \sum_{c_{\rm s},c_{\rm i}} \frac{ f(c_{\rm s},c_{\rm i})
  T_{\rm s}(c_{\rm s},n_{\rm s}) T_{\rm i}(c_{\rm i},n_{\rm i}) }{
  \sum_{n'_s,n'_i} T_{\rm s}(c_{\rm s},n'_{\rm s})
   T_{\rm i}(c_{\rm i},n'_{\rm i})
   p^{(l)}(n'_{\rm s},n'_{\rm i}) }.
\label{14}
\end{eqnarray}
The positive-operator-valued measures $ T_{\rm a} $, $ a ={\rm s,i} $,
characterize detection in the region with beam $ a $. We have for an iCCD
camera with $ N_a $ active pixels, detection efficiency $ \eta_a $ and mean
dark count number per pixel $ D_a $~\cite{PerinaJr2012}:
\begin{eqnarray}     
  T_a(c_a,n_a) &=& \left(\begin{array}{c} N_a \\ c_a \end{array}\right) (1-D_a)^{N_a}
   (1-\eta_a)^{n_a} (-1)^{c_a} \nonumber \\
  & &  \mbox{} \hspace{-18mm} \times  \sum_{l=0}^{c_a} \left(\begin{array}{c} c_a \\ l \end{array}\right)
    \frac{(-1)^l}{(1-D_a)^l}  \left( 1 + \frac{l}{N_a} \frac{\eta_a}{1-\eta_a}
   \right)^{n_a}.
\label{15}
\end{eqnarray}
Calibration of our iCCD camera \cite{PerinaJr2012a} gave us the following
parameters $ \eta_{\rm s}=0.230\pm 0.005 $, $ \eta_{\rm i}=0.220\pm 0.005 $,
$N_{\rm s}=N_{\rm i}=4096$, $ D_{\rm s}N_{\rm s} = D_{\rm i}N_{\rm i} =
0.040\pm 0.001$ for the signal (s) and idler (i) detection regions.

\section{Non-classicality and entanglement degradation caused by the increasing
noise}

\begin{figure} 
  \centerline{\includegraphics[width=0.95\hsize]{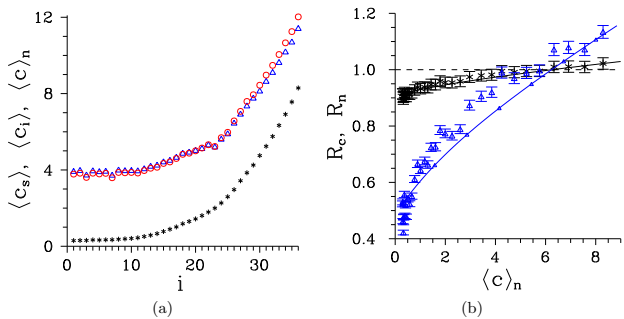}}
 \caption{(a) Mean experimental photocount numbers $ \langle c_{\rm s}\rangle $
  (blue $ \triangle $), $ \langle c_{\rm i}\rangle $ (red $ \circ $)
  and $ \langle c\rangle_{\rm n} $ ($ \ast $) in, in turn, signal-beam,
  idler-beam and noise detection region versus the number $ i $ identifying a TWB.
  (b) Noise-reduction-factors $ R_{\rm c} $ ($ \ast $) and $ R_{\rm n} $
  (blue $ \triangle $) determined for the experimental photocount histograms
  and reconstructed photon-number distributions of TWBs, respectively, as they
  depend on mean noise photocount number $ \langle c\rangle_{\rm n} $. Experimental data are plotted as isolated
  symbols with error bars derived from the number of measurement repetitions. Relative errors in (b) estimated from the
  data scattering are better than 3~\%. In (a), experimental errors are smaller than the plotted symbols. In (b),
  theoretical solid curves with appropriate symbols originate
  in the model, dashed line $ R = 1 $ indicates the non-classicality border.}
\label{fig2}
\end{figure}

To investigate degradation of the TWB entanglement as well as to analyze the
performance of the above entanglement quantifiers when the noise in the TWB
increases, the noise with multi-thermal photon statistics, originating in a
bulb lamp, was superimposed equally onto the signal and idler beams. An
increasing voltage applied to the bulb lamp leads to the increasing mean photon
numbers $ \langle n\rangle_{\rm n} $ of the noise field. 36 TWBs with different
levels of the noise were analyzed: Their mean photocount numbers $ \langle
c_{\rm s}\rangle $ and $ \langle c_{\rm i}\rangle $ in the signal and idler
detection regions, respectively, as well as the mean photocount numbers $
\langle c\rangle_{\rm n} $ of the noise field measured in the independent
detection are plotted in Fig.~\ref{fig2}(a).

We first roughly estimate the amount of non-classicality by applying the
noise-reduction-factor $ R $ \cite{Chesi2019} in Eq.~(\ref{1}) that, in fact,
quantifies the relative amount of paired photons in a TWB. The gradual decrease
of the relative amount of paired photons in the measured TWBs with the
increasing noise is monitored in Fig.~\ref{fig2}(b) by the increasing values of
the noise-reduction-factors $ R_{\rm c} $ and $ R_{\rm n} $ determined from the
photocount histograms and reconstructed photon-number distributions of the
analyzed TWBs, respectively. According to the graphs in Fig.~\ref{fig2}(b), the
TWBs with the mean noise photocount numbers $ \langle c\rangle_{\rm n} $
smaller than 5 are nonclassical ($ R_{\rm c}, R_{\rm n} < 1 $). As the
reconstruction algorithm qualitatively preserves the non-classicality while
improving it quantitatively, the curves for $ R_{\rm c} $ and $ R_{\rm n} $
mutually cross at $ R=1 $ where the transition to the classical region of $ R $
occurs.

The experimental results for the noisy TWBs are compared with the predictions
of the model that convolves the photocount (photon-number) distributions of the
independent noisy fields present in both the signal and idler beams with the
photocount histogram $ f^{\rm n-l} $ (photon-number distribution $ p^{\rm n-l}
$) of the original TWB without an additional noise using the formula analogous
to that in Eq.~(\ref{8}). The distributions of the noisy fields are given by
Eq.~(\ref{9}) in which we consider $ \langle c\rangle_{\rm n}$ ($ \langle
n\rangle_{\rm n} = \langle c\rangle_{\rm n} /\eta $) mean photocount numbers
(photon numbers) distributed into $ N_{\rm c} $ ($ N_{\rm n} $) equally
populated modes. Comparison with the experimental results suggests $ K_{\rm c}
= 110 $ independent modes in the noise fields to explain the loss of
non-classicality of the experimental photocount histograms $ f $. The slightly
smaller number $ K_{\rm n} = 90 $ of independent modes is appropriate in the
case of the reconstructed photon-number distributions $ p $. This is related to
the fact that the reconstruction with the positive-operator-valued measures $
T_{\rm a} $ in Eq.~(\ref{15}) partially reduces the noise.

\begin{figure} 
  \centerline{\includegraphics[width=0.95\hsize]{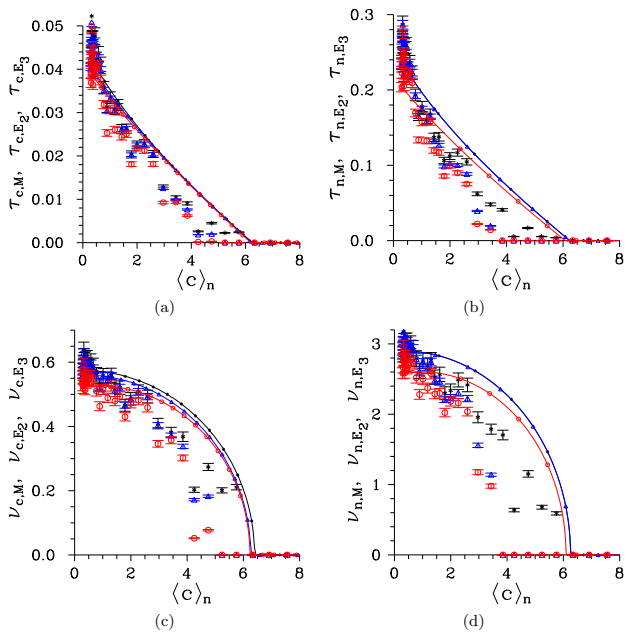}}
  \caption{Non-classicality depths $ \tau $ (a,b) and non-classicality counting parameters $ \nu $ (c,d) for NIs
  $ M $ (black $ \ast $), $ E_2 $ (blue $ \triangle $) and $ E_3 $ (red $ \circ $) for photocount histograms (a,c) and
  photon-number distributions (b,d) as they depend on mean noise photocount number $ \langle c\rangle_{\rm n} $.
  Experimental data are plotted as isolated symbols with error bars derived from the number of measurement repetitions,
  solid curves with appropriate symbols come from the model. Relative errors in (a,b) [(c,d)] estimated from the
  data scattering are better than 10~\% [5~\%].}
\label{fig3}
\end{figure}

The experimental as well as the theoretical values of both NDs $ \tau $ and
NCPs $ \nu $ drawn for different values of the mean noise photocount number $
\langle c\rangle_{\rm n} $ in Fig.~\ref{fig3} confirm the best performance of
the NI $ M $ in revealing the non-classicality of a whole multi-mode TWB. On
the other hand, the NI $ E_3 $ involving the third-order intensity moments
gives the worst results, in agreement with the findings of
Ref.~\cite{PerinaJr2017a}. Whereas the NI $ M $ identifies the non-classicality
of the TWB up to $ \langle c\rangle_{\rm n} \approx 6 $, the third-order
intensity moments of NI $ E_3 $ lose their ability to reveal the
non-classicality around $ \langle c\rangle_{\rm n} \approx 4 $. It is worth
noting that the commonly used noise-reduction factors $ R $ perform up to $
\langle c\rangle_{\rm n} \approx 5 $. The comparison of NCPs $ \nu $ drawn in
Figs.~\ref{fig3}(c,d) with the NDs $ \tau $ plotted in Figs.~\ref{fig3}(a,b)
shows comparable sensitivity of the NCPs in quantification of the
non-classicality from the point of view of the experimental errors under our
conditions. We note, however, that the NCPs cannot quantify the
non-classicality of highly quantum states~\cite{PerinaJr2019}. On the other
hand, the intensity moments do not have to be involved et all in the
determination of NCPs if the NIs based on the photocount (photon-number)
probabilities are applied \cite{PerinaJr2017a,PerinaJr2019}. In this case the
commutation relations, that depend on the number of field's modes, are not
needed. Substantial improvement of the amount of TWB non-classicality after the
reconstruction is evident when we compare the NDs $ \tau $ and NCPs $ \nu $
drawn in Figs.~\ref{fig3}(a,c) for the experimental photocount histograms $ f $
with those in Figs.~\ref{fig3}(b,d) appropriate for the reconstructed
photon-number distributions $ p $. The increase of non-classicality in the
reconstruction is due to partial elimination of the noise and, mainly,
correction for the finite detection efficiencies that brake the photon pairs
from which the non-classicality originates. The values of NDs $ \tau $ and NCPs
$ \nu $ are around 4-5 times larger after the reconstruction. This factor is
roughly proportional to $ 1/\eta $ which is a signature of the fact that the
mean photocount and photon numbers per one mode are smaller or comparable to 1.
For stronger fields, the mapping between the NDs $ \tau $ (NCPs $ \nu $)
belonging to the photocount histograms and the reconstructed photon-number
distributions is nonlinear (compare the condition $ \tau \le 1/2 $).

\begin{figure} 
  \centerline{\includegraphics[width=0.95\hsize]{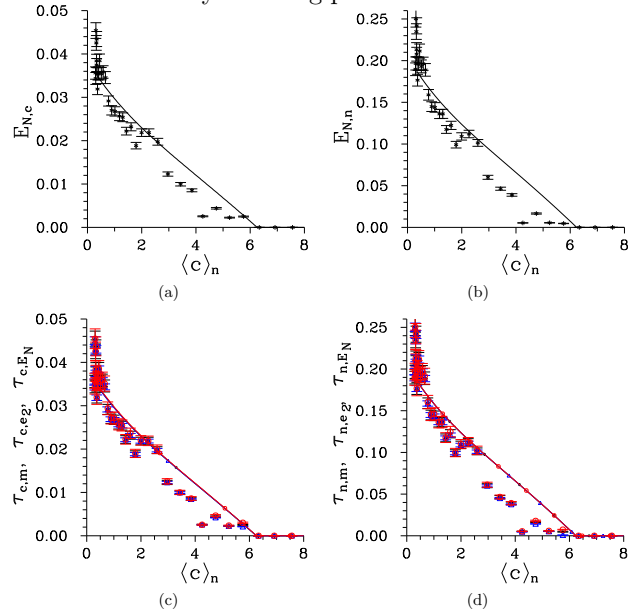}}
  \caption{Negativity $ E_{\rm N} $ (a,b) and non-classicality depth $ \tau $ (c,d) for NIs
  $ M $ (black $ \ast $), $ E_2 $ (blue $ \triangle $) and $ E_{\rm N} $ (red $ \circ $) for photocount histograms (a,c) and
  photon-number distributions (b,d) 'reduced' to a single-mode along Eq.~(\ref{12}) as they depend on mean noise photocount number $ \langle c\rangle_{\rm n} $.
  Experimental data are plotted as isolated symbols with error bars derived from the number of measurement repetitions,
  solid curves with appropriate symbols come from the model. Relative errors estimated from the
  data scattering are better than 10~\% for all plotted quantities.}
\label{fig4}
\end{figure}

The consideration of just one typical (average) mode of a TWB with its
intensity moments given along Eqs.~(\ref{11}) and (\ref{12}) leads to much
smaller values of the moments and thus the increased role of the noise.
Especially the odd-order moments are affected as the odd-order moments of
intensity fluctuations are sign-sensitive. We note that the measured TWBs were
composed of typically 50 modes determined by Eq.~(\ref{10}). In our case, this
disqualifies the use of third-order moments of the NI $ E_3 $ for
quantification of the non-classicality. On the other hand, the negativity $
E_{\rm N} $ determined from up-to the second-order intensity moments can
directly be used as an entanglement quantifier, as documented in
Figs.~\ref{fig4}(a,b). Alternatively, it can be considered as another NI and
then the corresponding NDs $ \tau_{E_{\rm N}} $ [see Figs.~\ref{fig4}(c,d)] and
NCPs $ \nu_{E_{\rm N}} $ can be calculated. In both cases, it identifies the
measured TWBs as entangled up to $ \langle c\rangle_{\rm n} \approx 6 $. The
comparison of NDs $ \tau_m $ and $ \tau_{e_2} $ [Figs.~\ref{fig4}(c,d)]
belonging to the NIs $ M $ and $ E_2 $ applied to single-mode moments with
those valid for the whole TWBs [Figs.~\ref{fig3}(a,b)] shows that the low-order
single-mode intensity moments successfully maintain the information about the
resistance of TWB non-classicality against the noise.

At the end, we note that the error bars plotted in the figures were determined
solely from the number of measurement repetitions. As such they do not reflect
instabilities and imperfections in the setup occurring during the measurements
of TWBs with different levels of the noise (one hour was typically needed to
characterize one TWB). Slow pump-beam intensity fluctuations, pump-beam
misalignment (temperature-induced position shifts) in the setup, temperature
stabilization of the iCCD camera and its synchronization with the laser source
were responsible for the main detrimental effects. The corresponding errors
were estimated from the experimental points in the graphs of Figs.~\ref{fig2},
\ref{fig3} and \ref{fig4}: Average relative errors were obtained by considering
all pairs of neighbor experimental points on a given experimental curve and
determining the mean value and the relative declination for each pair.

\section{Conclusions}

We have experimentally investigated deterioration of the entanglement of a twin
beam caused by an increasing external noise. We have suggested, verified and
mutually compared three experimentally feasible ways for quantifying the
twin-beam entanglement. The first two are based upon the non-classicality
depths and the non-classicality counting parameters of suitable
non-classicality identifiers. In the third way, the negativity is directly
determined for one typical mode of the TWB. The three entanglement quantifiers
perform comparably. They may be applied in any metrology, quantum-imaging or
quantum-information scheme that uses the twin beams and whose sensitivity to
the noise has to be quantified.

\section*{Acknowledgements} The authors thank I. I. Arkhipov for fruitful
discussions and suggestions. They acknowledge GA \v{C}R (project No. 18-08874S)
and M\v{S}MT \v{C}R (project No.~CZ.1.05/2.1.00/19.0377).


%

\end{document}